\documentclass{amsproc}

\usepackage{amssymb}

\usepackage{graphicx}

\usepackage[export]{adjustbox}
\usepackage{float}
\usepackage{url}

\newtheorem{theorem}{Theorem}[section]

\theoremstyle{definition}

\theoremstyle{remark}

\def\dc{/\!\!/}

\numberwithin{equation}{section}

\makeatletter
\def\moverlay{\mathpalette\mov@rlay}
\def\mov@rlay#1#2{\leavevmode\vtop{%
   \baselineskip\z@skip \lineskiplimit-\maxdimen
   \ialign{\hfil$\m@th#1##$\hfil\cr#2\crcr}}}
\newcommand{\charfusion}[3][\mathord]{
    #1{\ifx#1\mathop\vphantom{#2}\fi
        \mathpalette\mov@rlay{#2\cr#3}
      }
    \ifx#1\mathop\expandafter\displaylimits\fi}
\makeatother

\begin{document}

\title{\textbf{Conformal Hypergeometry and Integrability}}


\author[Volker Schomerus]{Volker Schomerus \\ \  \\ \textit{To Jasper Stokman on the occasion of his 50th birthday}}
\address{Deutsches Elektronen Synchroton DESY, Notkestrasse 85, 22607 Hamburg, Germany and
Universit\"at Hamburg, Luruper Chausee 149, 22761 Hamburg, Germany }
\curraddr{}
\email{volker.schomerus@desy.de}



\date{4/30/2021}

\begin{abstract}
Conformal field theories play a central role in modern theoretical physics with many
applications to the understanding of phase transitions, gauge theories and even the
quantum physics of gravity, through Maldacena's celebrated holographic duality. The
key analytic tool in the field is the so-called conformal partial wave expansion, i.e.
a Fourier-like decomposition of physical quantities into a basis of partial waves for
the conformal group SO$(1,d+1)$. This text provides an non-technical introduction to 
conformal field theory and the conformal bootstrap program with some focus on the 
mathematical aspects of conformal partial wave expansions. It emphasises profound 
relations with modern hypergeometry, group theory and integrable models of Gaudin 
and Calogero-Sutherland type.    
\end{abstract}

\maketitle

\section{Introduction}

During the last century physicists gradually learned to appreciate the importance of
conformal symmetry for our understanding of nature. By definition, conformal
symmetry is made up of angle preserving transformations. The most commonly
observed (conformal) symmetries are rotations and translations. These are
present in all homogeneous and isotropic systems. Beyond that, dilations
(scale transformations) are the most prominent examples of conformal
transformations. Scale invariance is an important phenomenon in physics
that refers to situations in which the features of a system do not change
upon zooming in or out. The first known occurrence in nature is the critical
point for the liquid-vapour transition in the phase diagram of water. Many
more of these points have been discovered in the meantime.  Some of them can
be realized in spin systems such as the famous Ising model.

Particle physicists started to appreciate conformal field theories
(CFTs) through the influential work of Wilson on the renormalization group
\cite{Wilson:1971bg}. By construction, scale invariance occurs at all fixed
points of renormalization group transformations. This gives CFTs a very special status
within the space of all theories, \footnote{Throughout this introduction I
will not distinguish clearly between scale and conformal symmetry. I will
briefly come back to the issue in section 2.} and makes them a good starting
point to explore, in particular, non-perturbative phenomena in quantum field
theory.
More recently, CFTs became the main actors on another stage, Juan Maldacena's
celebrated AdS/CFT duality \cite{Maldacena:1997re}. Its most general
formulation proposes an equivalence between non-gravitational quantum
systems in $d$ dimensions and gravitational systems in one dimension higher.
In this context, a
$d$-dimensional CFT is like a holographic screen for a gravitational theory
in Anti-deSitter (AdS) space. As we make progress in dealing with CFTs, the
AdS/CFT duality can be applied to provide novel views on the deep mysteries
of a quantum theory of gravity, such as the phenomenon of Hawking radiation
and black hole information.

The very brief history of scale/conformal invariance in physics that I
presented in the first two paragraphs may serve one main purpose: it explains
why a significant part of modern theoretical physics research is devoted to
CFTs and why the interest ranges over so many different communities, from
condensed matter theory and modern material science all the way to particle
physics and quantum gravity.

Invariance under conformal transformations turns out to be a surprisingly
powerful symmetry principle that strongly constrains physical observables
such as the correlations between different measurements. These constraints
may be exposed with the help of the most important analytical tool of CFT,
the so-called \textit{conformal partial wave} expansions. One should think
of these as an extension of Fourier analysis, but for systems with conformal
rather than mere translational symmetry. Just as plane waves in Fourier
analysis are uniquely characterized by their behaviour under translations,
conformal partial waves are determined by conformal symmetry alone, i.e.
they belong to the realm of group and representation theory.

When conformal field theories were considered for the first time about 50
years ago, the relevant group theory was very far from being developed, which
halted initial progress. This changed only after
physicists started to focus on $d=2$ dimensions where conformal symmetry becomes
infinite dimensional. For that case a very beautiful theory was developed
with many deep relations to mathematics and integrable systems. But this
is not our topic here. Instead, what I shall describe below is relevant
for CFTs in dimension $d > 2$. At the end of the last
century, and in particular with the discovery of the AdS/CFT duality, the
case for studying CFTs in $d > 2$ had become so strong that physicists
made another effort to develop at least some of its group theoretic
foundations. The influential work of Dolan and Osborn \cite{Dolan:2003hv,
Dolan:2011dv} on conformal partial waves for scalar 4-point functions and
some subsequent work by other authors, see in particular
\cite{Hogervorst:2013sma,Kos:2013tga,Kos:2014bka,Penedones:2015aga},
eventually provided a basis for some new breakthrough developments in
higher dimensional CFT, e.g. for the 3-dimensional critical Ising model.

At the time, physicists did not know that essentially all the required
results on conformal partial waves were already contained in the seminal
papers of Heckman and Opdam, see \cite{Heckman1987,Opdam1988,HeckmanBook}.
Somewhat surprisingly, before it was uncovered in \cite{Isachenkov:2016gim},
the profound relation went unnoticed between the study of integrable quantum
mechanics of Calogero-Moser-Sutherland type on the one hand and conformal
partial waves on the other. The main intent of this text is to help
reduce the apparent and unfortunate divide between the physics
communities that work on conformal field theories on the one hand and
mathematicians with expertise in modern group theory, harmonic analysis,
hypergeometry and integrability on the other. The study of CFTs continues
to challenge profound advances of its group theoretic ingredients. So it
is not too late for the relevant communities to join forces.
\medskip

After these introductory remarks, let me describe the plan of this short
note. The next section is meant as a smooth introduction to some basic
notions of critical systems and CFT. In the example of the critical Ising
model I will outline how statistical systems can become critical and
possibly conformal. In addition, I shall introduce the notions of
(scalar and spinning) fields, correlation functions, scaling weights,
conformal partial waves etc. and I will sketch the main idea of the
so-called conformal bootstrap program. Section 2 remains very descriptive,
without precise definitions. Nevertheless, I hope that it can build a first
intuition into the basic concepts of the area, in a spirit that is
relatively close to the physicists' approach. In section 3, I then rephrase
the basic setup in a more mathematical language. There, the notions of
fields, correlation functions and conformal partial waves are embedded
into group theory and harmonic analysis. Once this is achieved, we can
walk along well trodden paths into the world of hypergeometry and
integrable systems. We will see the first examples of the relation between
conformal partial waves and hypergeometry at the end of section 3. As I
will argue there, conformal partial waves for scalar 4-point functions
can be identified with the wave functions of hyperbolic Calogero-Moser-%
Sutherland models for the root system $BC_2$, i.e. these conformal
partial waves are (close relatives of) $BC_2$ Heckman-Opdam hypergeometric
functions. I conclude section 3 with a lightning overview of extensions
which are more or less well understood, or at least have been actively
researched already.

The final section is then devoted to a new important direction about
which very little seems to be known for now: multipoint correlation functions.
In a CFT, correlations of up to $N=3$ local measurements are essentially
determined by conformal symmetry. Hence, the first non-trivial correlations
that can be analysed with well studied conformal partial waves appears when
$N=4$. The territory we shall begin to explore in the final section lies just
one step beyond. For $N > 4$, conformal partial waves were largely unexplored
until very recently. Section 4 embeds the profound mathematical challenges in
gaining control of these objects into
the context of $N$-site Gaudin integrable models for the conformal group
\cite{Buric:2020dyz}. As I will discuss in detail, the study of
the relevant conformal partial waves requires us to consider certain limits of the
Gaudin integrable models which extend the so-called caterpillar or bending flow
limits previously considered in \cite{Chervov:2007dn,Chervov:2009,
rybnikov2016cactus}. In the final subsection I highlight an intriguing new
appearance of the lemniscatic elliptic Calogero-Moser-Sutherland model of
\cite{etingof2011107} in the context of $N$-point scalar conformal blocks
or, equivalently, in limits of $N$-site Gaudin integrable models.

While sections 2 and 3 of this note survey developments from several decades and do
not contain any new results, section 4 is based on some very recent research in
collaboration with I. Buric, S. Lacroix, J.A. Mann and L. Quintavalle
\cite{Buric:2021ywo,Buric:2021ttm}. 
\smallskip

\noindent
{\sc{Acknowledgments:}}
I wish to thank my co-authors Ilija Buric, Misha Isachenkov, Sylvain Lacroix, Pedro Liendo,
Yanick Linke, Jeremy Mann, Lorenzo Quintavalle, Evgeny Sobko for the inspiring collaboration
on the material presented in this note. At the same time I am very grateful to the organizers
of the workshop ``Hypergeometry, Integrability and Lie Theory'' at the Lorentz Center in Leiden 
for giving me the opportunity to present our work and to Jasper for providing the occasion. I 
have profited very much from many insightful questions, comments and discussions with 
participants during and after the meeting. This applies in particular to some of the new 
material included in the final subsection which has been stimulated by discussions with 
Pavel Etingof.

\section{Conformal Field Theory and Partial Waves}

\subsection{The critical Ising model}

As we have discussed in the introduction, conformal field theories are used to describe
the critical behavior of many statistical systems. One of the most well known
is the Ising model. Let us use this simple model to illustrate some of the key concepts.
The Ising model can be defined on a $d$-dimensional hypercubic lattice with lattice
sites given by $x_n = n^\mu \epsilon_\mu \in \mathbb{R}^d$. Here the summation over
$\mu = 1, \dots, d$ is understood and $\epsilon_\mu = \epsilon e_\mu$ coincide with
the unit vectors $e_\mu$ of $\mathbb{R}^d$ up to a common scale factor $\epsilon$.
We refer to $\epsilon$ as the lattice spacing. The coefficients $n^\mu$ are integers.

To each lattice site we now assign a spin $\sigma_n \in \{\pm 1\}$. Any such choice
of spin assignments on all lattice sites is referred to as a spin configuration. The
object $\sigma$ is (a discrete version) of what physicists call a \textit{field}. In
order to define our theory we must specify a measure on the on the space of
configurations. For the Ising model, this measure is
\begin{equation} \label{Isingmeasure}
P_\zeta(\{\sigma\}) = \mathcal{N} e^{\zeta \sum_{\langle n m\rangle} \sigma_n \sigma_m }\ .
\end{equation}
Here the sum extends over nearest neighbors $\langle n m \rangle$, i.e. over all
pairs of lattice sites that are a distance $|x_n - x_m| = \epsilon$ apart.

The nature of a typical configuration selected by the measure \eqref{Isingmeasure} depends
very much on the value of the parameter $\zeta$. If we set $\zeta=0$, for example, neighboring
spins do not interact, i.e.\ the value the measure assumes is the same on all spin configurations.
Hence, if we measure the spin at site $n$, this won't tell us anything about the value of the
spin at another site $m$, in other words there is no correlation between the value $\sigma_n$ of
the spin at site $n$ and the value of $\sigma_m$ at another sites $m \neq n$ of the lattice. In
the opposite case with $\zeta = \infty$, configurations in which spins assume different values
are infinitely suppressed, i.e.\ the measure only allows for the two constant configurations
$\sigma_n = 1$ or $\sigma_n = -1$ for all sites $n$. In this case, the correlations between
spins are as strong as they can be. We only need to measure the spin $\sigma_n$ at one site
$n$ to know its value at all the other sites $m \neq n$. Once we go away from this extreme
case, spin flips are possible but still costly. In other words, while the measure still
assumes its maximum value on the two constant configurations, it becomes less strongly peaked
as we decrease $\zeta$. Consequently, spin configurations with large `islands' of aligned spins
will start to play a role. As we reduce the value of $\zeta$, smaller islands contribute. It
turns out that, for some very special value $\zeta = \zeta_c$, islands of all sizes appear
with equal probability. At this very special point, the system appears to be self-similar on
scales that are large compared to the lattice spacing $\epsilon$ (and small compared to the
overall size $L$ of the lattice if we consider a finite lattice). It is this `critical' value
$\zeta = \zeta_c$ of the parameter $\zeta$ that we want to focus on.

In order to probe the Ising model, physicists measure correlations between spins. The simplest
interesting example is the correlation between two spins $\sigma_n$ and $\sigma_m$ at two
different lattice sites $n,m$. The quantity is defined as
\begin{equation} \label{eq:2ptsum}
\langle \sigma_n \sigma_m \rangle_\zeta = \sum_{\{\sigma\}} P_{\zeta}(\{\sigma\})
\, \sigma_n \sigma_m \ .
\end{equation}
To make this sum well defined, one can imagine that it is carried out on a finite lattice
with $L^d$ lattice sites, even though criticality requires $L$ being sent to infinity at the
end. As $\zeta$ approaches $\zeta_c$ the system becomes self-similar in the sense described above
and hence this correlation function must take the form
\begin{equation} \label{eq:2ptfct}
\langle \sigma_n \sigma_m \rangle_{\zeta_c} \sim \frac{1}{|x_n-x_m|^{2\Delta_\sigma}}\ .
\end{equation}
Here, $|x_n-x_m|$ denotes the Euclidean distance between the lattice points $x_n$ and $x_m$
in $\mathbb{R}^d$. The $\sim$ means the function on the right hand side approximates the
correlation function on intermediate scales or, equivalently, in the limit of an infinite
lattice size $L$ and vanishing lattice spacing $\epsilon$. The parameter $\Delta_\sigma$
that describes how the correlation function falls off as we increase the distance between
the two lattice sites is referred to as the \textit{scaling weight} of $\sigma$. It is a
very important observable of the critical system.

Determining the value of $\Delta_\sigma$ is one of the central challenges of CFT. In $d=2$
dimensions, it is possible to give an exact value. As was observed
in \cite{Belavin:1984vu}, the scaling weight of the spin field in the critical 2-dimensional
Ising model can be calculated using representation theory of the Virasoro algebra with central
charge $c=1/2$. This approach famously gives
\begin{equation}
\Delta^{\textrm{[2D]}}_\sigma = 1/8\ .
\end{equation}
For the 3-dimensional critical Ising model, the exact value of the scaling weight
$\Delta_\sigma$ is not known, but there exist good numerical simulations (NSs) of
the correlation \eqref{eq:2ptsum} near $\zeta = \zeta_c$ that give the approximate
result \cite{Hasenbusch:2010hkh}
\begin{equation}
\Delta^{\textrm{[3D],NS}}_\sigma \sim 0.51813(5)\ .
\end{equation}
Let us point out right away that a more precise value of this exponent is known
by now. However, this precision determination is based not on numerical simulation
of the Ising model, but rather on the conformal bootstrap program that we shall
outline below.

Before we go there, let us briefly mention that the scaling weight $\Delta_
\sigma$ of the spin field $\sigma$ is not the only field in the Ising model.
Obviously, it is possible to construct many similar quantities. To give just one
additional example, let us introduce the object
\begin{equation}
\varphi^\mu_n = \sigma(x_n) \nabla^\mu \sigma(x_n) = \sigma(x_n)
\sigma (x_n + \epsilon_\mu) - 1 \ .
\end{equation}
Note that the definition of $\varphi^\mu$ depends on the choice of a direction
$\mu$ in the $d$-dimensional space. Physicists refer to $\varphi^\mu$ as a vector
field as opposed to the spin field $\sigma$ which is scalar, i.e.\ does not single
out any direction. One can go further and consider fields that are sensitive to
the value of the Ising spin at more than two sites. The relevant sites may fall
into a local cluster or be located along lines, surfaces, etc. One calls the
associated objects local, line, surface operators, respectively. These operators
can be inserted into correlation functions of the form \eqref{eq:2ptsum}, just
as we did for the spin field $\sigma$. It turns out that the 2-point correlations
of local operators also fall off with some characteristic exponent $\Delta_\varphi$
whose value depends on the field $\varphi$ that is inserted.

\subsection{Conformal field theory}

Critical systems are included in the wider class of homogeneous and isotropic
systems which are invariant under translations $x^\mu \mapsto x^\mu + a^\mu$
and rotations $x^\mu \mapsto O^\mu_{\, \nu} x^\nu$ of points $x = (x^\mu) \in
\mathbb{R}^d$. As we have described above, the characteristic feature of
critical system is self-similarity, i.e.\ at the critical point the symmetry
is enhanced to include scale transformations $x^\mu \mapsto \lambda x^\mu$
with $\lambda \in \mathbb{R}^+$. It turns out, however, that the symmetry of
many critical systems, including the critical Ising model, is further enhanced
to the $d$-dimensional conformal group $SO(1,d+1)$. In addition to translations,
rotations and scale transformations, the conformal group possesses $d$ generators
of so-called special conformal transformations. This enhancement has dramatic
consequences as we shall discuss now.

To begin exploring these consequences let us point out that in a conformal field
theory, correlation functions of three fields are fixed by symmetry, up to some
overall constant. This follows from the well known fact that conformal
transformation can be used to map any set of three points in $\mathbb{R}^d$
to the values $x_1 = 0, x_2 = \vec{e}_1, x_3 = \infty$. Hence, a 3-point
function of scalar fields $\phi_i$ of scaling weight $\Delta_i$ must take
the form
\begin{equation} \label{eq:3point}
\langle \phi_1(x_1) \phi_2(x_2) \phi_3(x_3) \rangle =
\frac{\gamma_{123}}{|x_{12}|^{\Delta_{12,3}} |x_{23}|^{\Delta_{23,1}}
|x_{13}|^{\Delta_{13,2}}} = \gamma_{123}
\includegraphics[width=2cm,valign=c]{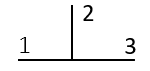}
\end{equation}
where $x_{ij} = x_i-x_j$ and $\Delta_{12,3} = \Delta_1 + \Delta_2 - \Delta_3$
etc. In more mathematical terms, the $x_i$-dependent function on the right hand
side may be thought of as Clebsch-Gordon coefficients for the conformal group
and the prefactors $\gamma_{123}$ provide a multiplication table for scalar
fields of the theory. This is the interpretation we have tried to stress in
the symbolic representation of the 3-point function on the right hand side.

What is so special about conformal field theories is indeed that their
`multiplication table' is really just a set of constants. To avoid confusion
we need to stress that the `multiplication' does not close on scalar fields,
i.e. 3-point functions of two scalars and one vector field, for example, can
be non-vanishing. But this does not alter the basic conclusion that conformal
symmetry guarantees that there exists a `multiplication table' of constants.
It was already observed in the early days of CFT that the
multiplication table of conformal field theories must obey stringent
consistency conditions that are reminiscent of associativity constraints
on multiplication tables in algebra. This observation is the main content
of Polyakov's conformal bootstrap program \cite{Polyakov:1974gs}.
The concrete implementation of this program, however, is rather
difficult and so far analytic solutions were only found for 2-dimensional
conformal field theories. Nevertheless, the modern conformal bootstrap has
led to profound new insights on conformal field theories in $d \geq 3$, see
below.

\subsection{Conformal partial waves and the bootstrap}

The most basic associativity constraints arise from the study of 4-point
correlation functions. These are no longer determined by conformal symmetry
alone simply because one can form two independent conformally invariant
cross ratios from four points in $\mathbb{R}^d$. We shall parametrize these
through two variables $u_1$ and $u_2$. In order to analyse 4-point functions
it is useful to expand them in terms of a `basis' of conformal partial waves
\begin{equation} \label{eq:CPWexp}
\langle \prod_{i=1}^4 \phi_i(x_i) \rangle \sum_{\Delta,l}
\gamma_{12;\Delta,l} \gamma_{34;\Delta,l} G_{\Delta,l}(u_1,u_2)
\end{equation}
Here the sum extends over all primary (= non-derivative) fields of scaling
weight $\Delta$ and spin $l$, i.e. over fields that possess $l$ symmetric
vector indices $\mu_1,\dots, \mu_l$. Note that the coefficients $\gamma$
are part of the multiplication table, though they also include multiplication
of scalar fields into fields that carry spin, unlike the coefficients we
introduced in eq.\ \eqref{eq:3point}. Generalizing the graphical notation
we introduced in eq.\ \eqref{eq:3point} we can write such 3-point functions
as
\begin{equation}
\langle \phi_1(x_1) \phi_2(x_2) \Phi_3(x_3) \rangle = \gamma_{12;\Delta,l}
\includegraphics[width=2.3cm,valign=c]{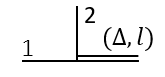} \ .
\end{equation}
The Clebsch-Gordan coefficients for two scalar and one spinning field are
certainly known explicitly. But since we won't need such detail, we shall
content ourselves with the graphical notation. The conformal partial waves
that appear in the expansion \eqref{eq:CPWexp} of the scalar 4-point
function can be considered as 4-point Clebsch-Gordan maps of the form
\begin{equation}
G_{\Delta,l} = \includegraphics[width=2.5cm,valign=c]{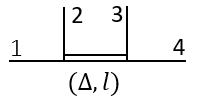}\ .
\end{equation}
The graphical notation in this equation emphases the construction of the
partial wave by contraction of two (3-point) Clebsch-Gordan coefficients.
One can actually translate this into a precise integral formula for $G$,
\begin{equation} \label{eq:int}
G_{\Delta,l} \sim \int d^dx_0 x_{10}^{l+a-\Delta} x_{20}^{l-a-\Delta}
x_{30}^{l - b + \Delta - d} x_{40}^{l_b+\Delta-d} \left(|\mu||\tilde \mu|\right)^l
Y^d_l\left(\frac{\mu\cdot\tilde \mu}{|\mu||\tilde \mu|}\right)
\end{equation}
where
\begin{equation}
\mu = \frac{x_{10}}{x_{10}^2} - \frac{x_{20}}{x_{20}^2} \quad , \quad
\tilde \mu = \frac{x_{40}}{x_{40}^2} - \frac{x_{30}}{x_{30}^2},
\end{equation}
the parameters $a,b$ are determined by the weights $\Delta_i$ as $2a = \Delta_2 -
\Delta_1$, $2b = \Delta_3- \Delta_4$ and $Y^d_l$ denotes zonal spherical functions.
We have also used the notation $x^t = |x|^t$ whenever elements $x \in \mathbb{R}^d$
are raised to some power $t$. The integral over $x_0 \in \mathbb{R}^d$ takes care of
the sum over intermediate states when we construct a 4-point invariant from two 3-point
invariants. Since the relevant representations of the conformal group are infinite
dimensional (see next section), the summation is most easily expressed as an integral
over some insertion point $x_0$ of the intermediate field of weight $\Delta$ and spin
$l$.

Before we discuss which special functions the integrals \eqref{eq:int} integrate
to we want to sketch one important application of conformal partial wave expansions
that should suffice to illustrate their importance. The attentive reader might have
noticed already that the expansion \eqref{eq:CPWexp} breaks the symmetry of the
correlation functions with respect to the permutation of the four scalar fields.
The expansion we have displayed pairs the fields $\phi_1$ and $\phi_2$ together.
This arbitrary choice is referred to as the choice of a \textit{channel}. To match
common convention, we shall use the term operator product expansion (OPE) channel
from now on, though it would be more appropriate to call it a conformal partial
wave (CPW) channel here. In any case, for $N=4$ scalar fields there are two more
channels, in addition to the one we have displayed, for which $\phi_1$ is paired
with either $\phi_3$ or $\phi_4$. Note that the conformal partial waves $G$ do
depend on the choice of the OPE channel. But after summing the partial waves
with the coefficients $\gamma$, the symmetry between the four fields should be
restored. This provides an important consistency condition on the coefficients
$\gamma$ that is known as crossing symmetry condition. The conformal bootstrap
\cite{Polyakov:1974gs} is designed to exploit these constraints in order to
compute observables of the theory, such as the values of the scaling weights
$\Delta$ or the coefficients $\gamma$. The program has had incredible success
in $d=2$ dimensions after the seminal paper of Belavin, Polyakov and Zamolodchikov
\cite{Belavin:1984vu}. During the last decade, the conformal bootstrap has seen
an impressive revival in higher dimensions with new numerical as well as analytical
incarnations, see \cite{Rattazzi:2008pe,Rattazzi:2010yc, Poland:2011ey,ElShowk:2012ht,
El-Showk:2014dwa} as well as \cite{Fitzpatrick:2012yx,Komargodski:2012ek,
Alday:2015eya,Alday:2015ota,Kaviraj:2015cxa,Kaviraj:2015xsa} for early contributions
and \cite{Poland:2016chs,Simmons-Duffin:2016gjk,Poland:2018epd} for reviews, though
geared to physicists. This has produced many stunning new insights including record
precision computations of critical exponents in the critical Ising model
\cite{Kos:2016ysd} through the numerical conformal bootstrap (nCB),
\begin{equation}
\Delta^{\textrm{[3D],nCB}}_\sigma \sim 0.5181489(10)\ .
\end{equation}
Note that the bootstrap analysis succeeded to increase the precision to which we
know the scaling weight $\Delta_\sigma$ of the spin field of the Ising model in
$d=3$ dimensions by two orders of magnitude, as compared to all the clever
numerical simulations that had been developed over a long time. This celebrated
result has triggered a tremendous amount of new research that uses the bootstrap
program to compute observables in (super)conformal field theories in higher
dimensions.

The determination of $\Delta_\sigma$ for the $d=3$ critical Ising model required
very good knowledge of the conformal partial waves which need to be evaluated
in a power series expansion around one special point $u_1^\ast, u_2^*$ in the
space of cross ratios to very high orders. When conformal partial wave expansions
were first introduced in \cite{Ferrara:1972uq,Ferrara:1973vz} the partial waves
remained elusive objects. The necessary mathematical results that made efficient
evaluation of the partial waves possible were pretty much out of reach back then.
Physicists wondered for more than three decades what kind of special functions
conformal partial are in dimension $d > 2$.\footnote{In $d=2$ dimensions it is
easy to perform the integrals and to express them as a sum of products of Gauss'
hypergeometric functions.} They were interested to understand how partial waves
depend on the cross ratios $u_i$, the parameters $\Delta, l$ of the `intermediate
field' and on the weights $\Delta_i$ of the external fields. When higher
dimensional conformal field theories became
a central research topic at the end of the last century, Dolan and Osborn
found a way to address these questions. Their idea was to characterize the partial
waves as eigenfunctions of Casimir differential operators \cite{Dolan:2003hv}. By
analysing these operators they were able to uncover many interesting properties
and in particular to express conformal partial waves in even dimensions through
sums of products of Gauss' hypergeometric functions \cite{Dolan:2003hv,Dolan:2011dv}.
After a number of relevant extensions \cite{Hogervorst:2013sma,Kos:2013tga,Kos:2014bka,
Penedones:2015aga}, this line of research provided enough control over the conformal
partial waves to implement a numerical version of Polyakov's bootstrap program.

The central observation of \cite{Isachenkov:2016gim} implied later that all the results
physicists had obtained on conformal partial waves for scalar 4-point functions must be
contained in the seminal series of papers by Heckman and Opdam \cite{Heckman:1987a,
Heckman:1987b,Opdam1988a,Opdam1988b}, see also \cite{HeckmanBook,OpdamDunkl}. From the
vantage point of \cite{Isachenkov:2017qgn} it became clear that  the conformal partial
waves \eqref{eq:int} are sums of Harish-Chandra functions associated with the root
system $BC_2$ and very close relatives of the corresponding Heckman-Opdam hypergeometric
functions. At this point we will leave the more physics oriented approach of the first
section to explain the relation between conformal partial waves and hypergeometry in a
more mathematical language that should be easier to follow than the original work on
the subject.

\section{Conformal Partial Waves and Hypergeometry}

\subsection{Group theoretic reformulation}

In order to understand the natural home of conformal partial waves in hypergeometry,
we now want to start translating the physics ingredients we described above
into a more mathematical language, mostly in the context of group theory. Recall that the
symmetry group of a CFT is $G = SO(1,d+1)$ (or some covering thereof).
The fields we worked with in the previous sections were inserted at points $x$ in the
$d$-dimensional space. Since the action of conformal transformations is transitive, it is
clear that we can think of space as a quotient of the conformal group by the stabilizer
of a single point. The latter is the parabolic subgroup $P$ that is generated by dilations,
rotations and special conformal transformations. Dilations form a subgroup $SO(1,1) \subset G$
while rotations are elements of $SO(d) \subset G$. These act on $\mathbb{R}^d$ in the usual way
and hence they act on the $d$-dimensional abelian subgroup $N \subset G$ that is generated by
special conformal transformations. Altogether, the stabilizer $P$ of a single point takes the
form $P = K \ltimes N$ where $K = SO(1,1) \times SO(d)$. Note that $K$ is not compact. From
now on we shall think of a point $x$ as an element $x \in G /P$ in the quotient space of the
conformal group $G$ and the parabolic subgroup $P$.

The next ingredient we need to transfer to mathematics is the notion of a field. Configurations
of a scalar field $\phi$ of weight $\Delta$ are elements in the space $\Gamma_\Delta$ of
sections of a line bundle over the quotient space $G/P$. More precisely, we define
\begin{equation} \label{eq:GammaGP}
\Gamma_\Delta = \Gamma_\Delta(G/P) = \{ f: G \rightarrow \mathbb{C} \, | \,
f(gp) = \chi_\Delta(p^{-1}) f(g) \}  \ .
\end{equation}
The object $\chi_\Delta$ is a character of the parabolic subgroup $P \subset G$ that is trivial
on the subgroups $N$ and $R = SO(d)$. On the generator $D$ of scale transformations, on the other
hand, the character $\chi_\Delta$ assumes the value $\chi_\Delta(D) = \Delta$. Note that $D$ is an
element of the conformal algebra $\mathfrak{g} = \mathfrak{so}(1,d+1)$ and we did not
distinguish in notation between a character of the conformal group and its Lie algebra. Let us
also stress that the linear spaces $\Gamma_\Delta$ of sections come equipped with the left regular
action of the conformal group. For generic values of $\Delta$ the representation of $G$ on
$\Gamma_\Delta$ is irreducible and infinite dimensional.

After this preparation we can now state that an $N$-point correlation function may be
considered as a conformally invariant element in the tensor product of $N$ spaces
$\Gamma_1, \dots, \Gamma_N$,
\begin{equation}
\langle \varphi_1(x_1) \cdots \varphi_N(x_N) \rangle \in
\left(\Gamma_1 \otimes \cdots \otimes \Gamma_N \right)^G
\end{equation}
where the superscript $G$ is used to denote the subspace of $G$-invariant elements. To
be more precise, the correlation function itself is not invariant, but there exists a
map from the correlation function to the associated invariant. The latter is mediated
by multiplication with a certain function $\Omega(x_i;\Delta_i)$ of the insertion points
$x_i$. While $\Omega$ is not unique, there exists a canonical choice \cite{Buric:2019dfk}.

\subsection{Conformal partial waves as spherical functions}

For the moment let us focus on the case with $N=4$ scalar field insertions. The space of
4-point conformal invariants can be evaluated in two steps. First we want to evaluate
the tensor product of the tensor factors $\Gamma_1 = \Gamma_{\Delta_1}$ and $\Gamma_2 =
\Gamma_{\Delta_2}$. The result we are going to spell out in a moment involves the
following linear space
\begin{equation} \label{eq:GammaGK}
\Gamma_{\Delta} (G/K) \cong  \{ f: G \rightarrow \mathbb{C} \, | \,
f(gk) = \chi_\Delta(k^{-1}) f(g) \} \ .
\end{equation}
The definition is similar to the construction in eq.\ \eqref{eq:GammaGP}, except that the
right covariance condition only involves elements of the subgroup $K \subset P \subset G$
of the conformal group. In an abuse of notation we have denoted the character of this
subgroup $K$ by the same symbol $\chi_\Delta$ as in the case of the parabolic subgroup
$P$ above. Note that the two sets of characters are related by restriction since we
assumed the character of $P$ to be trivial on the abelian subgroup $N \subset P$.
With this notation we can now state \cite{Dobrev:1977qv}
\begin{equation}
\Gamma_i(G/P) \otimes \Gamma_j(G/P) \cong \Gamma_{\Delta_{ij}}(G/K)
\quad
\end{equation}
where $\Delta_{ij} = \Delta_i - \Delta_j$. We can put this formula to use in
the following analysis of conformal invariants in the 4-fold tensor product,
\begin{eqnarray*}
\left( \Gamma_{\Delta_1} \otimes \Gamma_{\Delta_2} \otimes \Gamma_{\Delta_3} \otimes
\Gamma_{\Delta_4} \right)^G & \cong &  \left(\Gamma_{\Delta_{12}}(G/K) \otimes
\Gamma_{\Delta_{34}}(G/K)\right)^G  \noindent \\[2mm]
& \cong & \Gamma_{\Delta_{12}, \Delta_{43}} (K \backslash G / K)\ .
\end{eqnarray*}
The final result is expressed in terms of the spaces
\begin{equation}
\Gamma_{\Delta,\Delta'}(K\backslash G /K) =  \{ f: G \rightarrow \mathbb{C} \, | \,
f(k_lgk_r) = \chi_{\Delta_{21}}(k_l) \chi_{\Delta_{34}}(k_r^{-1}) f(g) \} \ .
\end{equation}
These consist of sections of a line bundle over the double coset space
$G\dc K$. It is not difficult to see that the left action of $K$ in $G/K$ is
transitive with stabilizer subgroup $B = SO(d-2)$. It follows that the double
coset space has dimension
$$ \textit{dim}\, (G\dc K) = \textit{dim}(G) - 2 \textit{dim}(K) + \textit{dim}(B) = 2\ . $$
The double quotient $G\dc K$ coincides with the space of conformally invariant cross
ratios that can be build from $N=4$ points in $\mathbb{R}^d$ for $d \geq 2$.

\subsection{Hypergeometry of conformal partial waves}

It is obvious that the Laplace operator on the conformal group $G$ descends to the
space $\Gamma(K\backslash G/K)$. The associated radial Laplacian has been constructed
and studied extensively in the mathematical literature \cite{HarishChandra,HelgasonBook}.
In particular, it is well known \cite{Olshanetsky:1981dk,Olshanetsky:1983wh} to be
intimately related the trigonometric/hyperbolic Calogero-Moser-Sutherland Hamiltonian
\cite{Calogero:1970nt,Sutherland:1971ks,Moser:1975qp}
\begin{eqnarray}
  H_{\textrm{CS}} &=& - \sum_{i=1}^{2} \frac{\partial^2}{\partial u_i^2} +
  \frac{k_m(k_m-1)}{2}\left[\sinh^{-2}\left(\frac{u_1+u_2}{2} \right)
  + \sinh^{-2}\left(\frac{u_1-u_2}{2}\right) \right]\nonumber \\[2mm]
   & & \hspace*{1cm} + \sum_{i=1}^{2}  \left[ k_l(k_l-1) \sinh^{-2}(u_i) +
   \frac{k_s(k_s + 2 k_l -1)}{4} \sinh^{-2} \left(\frac{u_i}{2}\right) \right]
\end{eqnarray}
associated to the roots system $BC_2$ with multiplicity parameters given by
\begin{equation}\label{eq:multiplicities}
k_s = \Delta_{43} \ , \quad k_m = \frac12 ( \Delta_{21} - \Delta_{43} + 1) \ ,
\quad k_l = \frac12 (d-2)
\end{equation}
for the short, middle and long roots, respectively. These Hamiltonians have also
been discussed in the talks of Margit R\"ossler and Eric Opdam at the meeting.
The line of reasoning we have outlined here now allows to draw an important conclusion:
In the case of scalar 4-point functions, the sought after conformal partial waves are
eigenfunctions of a Calogero-Moser-Sutherland Hamiltonian and hence they are closely
related with Heckman-Opdam hypergeometric functions for the $BC_2$ root system. To be
quite precise, the boundary conditions one has to impose in the case of conformal
partial waves differ from those imposed in the work of Heckman and Opdam. The correct
conditions are described in \cite{Isachenkov:2017qgn}. It turns out that conformal
partial waves are a linear combination of four of the eight Harish-Chandra functions
in the problem.

In spite of this minor difference, the relation with Heckman-Opdam hypergeometry
can be used to deduce many important properties in conformal partial waves, see
\cite{Isachenkov:2017qgn}. The position of poles and their residues can be obtained
easily from those of the corresponding Harish-Chandra functions, see e.g.
\cite{HeckmanBook,OpdamDunkl}. To obtain series expansions we realized the partial
waves as the $q \mapsto 1^-$ limit of virtual Koornwinder polynomials \cite{Rains:2005}
and employed the binomial expansions of Koornwinder polynomials from \cite{Okounkov1998}
to perform the degeneration with the help of \cite{Stokman2005LimitTF,Koornwinder2014},
see appendix A of \cite{Isachenkov:2017qgn} for details. Interesting shift equations
for the various parameters of partial waves, finally, arise from the bispectral
duality \cite{Ruijsenaars:1988pv} between the trigonometric Calogero-Moser-Sutherland
model with the rational Ruijsenaars-Schneider model for the same root system
\cite{Ruijsenaars:1986vq}. In this way, all properties of conformal partial waves
physicists needed to prepare for applications in the modern conformal bootstrap can
be recovered with relatively little effort from results mathematicians had obtained
at the end of the last century.

The relation between conformal partial waves and Calogero-Moser-Sutherland models was
first observed in \cite{Isachenkov:2016gim} based on a direct analysis of the Casimir
operators in the work of Dolan and Osborn. The direct group theoretical explanation I
have outlined in the previous section has been described in \cite{Schomerus:2016epl,
Schomerus:2017eny}, with inspiration from earlier work in particular by Feher
\cite{Feher:2009wp}.
\medskip

This group theoretic approach to conformal partial waves admits many
relevant extensions. Among them is the extension to correlation functions of
fields with spin, i.e.\ fields that transform non-trivially under the action
of the rotation group $SO(d) \subset G$. In the general setup we described,
this extension can be implemented by replacing the characters $\chi_\Delta$
of the parabolic subgroup $P$ or its subgroup $K = SO(1,1) \times SO(d)$ in
eqs.\ \eqref{eq:GammaGP} and \eqref{eq:GammaGK}, respectively, by some finite
dimensional representations. The radial Laplacians on the
associated vector bundles over the double coset $G\dc K$ have been worked out
very explicitly \cite{Schomerus:2016epl,Schomerus:2017eny} for a few
representations of the rotation group $SO(d)$ in $d=3,4$. A general theory of
the Hamiltonians and their eigenfunctions remains to be developed, even though
for applications in physics sufficient control has been achieved, see
\cite{Echeverri:2016dun,Karateev:2017jgd} and further references therein.

Another very important extension is required to study superconformal field
theories, i.e.\ quantum field theories in which the conformal Lie algebra
$\mathfrak{g} = \mathfrak{so}(1,d+1)$ gets extended to a superconformal one.
This is particularly important since most known conformal quantum field theories
are actually supersymmetric, especially in dimension $d > 3$. For superconformal
algebras of type I a general theory is obtained in \cite{Buric:2019rms,Buric:2020buk}.
It reduces the theory of superconformal partial waves to that of ordinary conformal
partial waves for fields with spin, see previous paragraph. The CFT literature
contains few results in superconformal partial waves, except for very exceptional
choices of field insertions, see e.g.\ \cite{Buric:2019rms} for references to the
physics literature.

There is one more extension we want to briefly mention before we conclude this
section. So far we have mostly discussed correlation functions of fields $\varphi(x)$
that are inserted at a point $x$ in the $d$-dimensional Euclidean space $x \in
\mathbb{R}^d$ or its conformal compactification. But physics applications require to
also consider operators that are localized along higher dimensional geometries such
as lines, surfaces etc. One such example are the famous Wilson line operators in
conformal gauge theories. Such non-local operators (fields) can be treated within
the same framework, except that the stabilizer group $P$ of a point $x$ gets replaces
by the stabilizer group $P_p = SO(1,p+1) \times SO(d-p) \ltimes \mathbb{R}^d$
of some $p$-dimensional submanifold. In this context, 2-point correlation functions
of non-local fields can be treated pretty much in the same way as the 4-point function
of local fields.\footnote{A non-local field that is inserted along a submanifold of
dimension $p=0$ may be considered as a `compound' of two local fields.} The theory of
conformal partial waves for 2-point correlation functions of non-local fields in CFT
has been developed systematically in \cite{Isachenkov:2018pef}. Not surprisingly, it
can be reduced to the study of the Laplace operator on double quotients of the form
$K_p \backslash G / K_{p'}$ with $K_p = SO(1,p+1) \times SO(d-p)$. Therefore it is
related to Heckman-Opdam hypergeometric functions for root systems $BC_n$ with the
rank $n$ given by
\begin{equation}\label{rank}
n = \textit{min} \left(p+2,p'+2,d-p,d-p'\right)\ .
\end{equation}
Note that the rank is $n=2$ in case at least one of the non-local fields is
inserted along a surface of dimension $p = 0$ and $d\geq 2$. Higher ranks appear
e.g. for correlation functions of two line operators (i.e.\ $p = p' = 1$) in
$4$-dimensional conformal field theories etc. The theory of the associated
conformal partial waves is relatively well developed, but this time only due
to its relations with $BC_n$ Heckman-Opdam theory \cite{Isachenkov:2018pef}.

\section{Integrability of Multipoint Conformal Partial Waves}
\def\a{\alpha}
\def\H{\mathcal{H}}
\def\Ht{\tilde{\mathcal{H}}}
\def \Dg {\Delta}
\def \dg {\delta}
\def \ds {\partial}
\def \ag {\alpha}
\def \bg {\beta}
\def \cg {\gamma}
\newcommand{\cupdot}{\charfusion[\mathbin]{\cup}{\cdot}}
\def\lieg{\mathfrak{g}}
\def\lieh{\mathfrak{h}}
\def\liem{\mathfrak{m}}
\def\liep{\mathfrak{p}}
\def\liek{\mathfrak{k}}
\def\id{\textit{id}}
\def\Spin{\textit{Spin}}
\def\g{\mathfrak{g}}
\def\Lc{\mathcal{L}}
\def\Hc{\mathcal{H}}
\def\cG{\mathcal{G}}
\def\dep{\mathfrak{d}}
\def \om {\omega}

Now that have described the link between conformal partial waves and hypergeometry
at the example of scalar 4-point functions and have sketched a few extensions at the
end of the previous section, I want to dedicate the final section to one challenging
new direction that is only beginning develop. In the study of concrete conformal field
theories physicists often compute correlation functions of more than $N=4$ fields. And
even though the conformal bootstrap in its original formulation is focused on 4-point
functions, one could very well imagine an alternative program that is based multipoint
partial waves. Note that associativity of an algebra can be studied either within the
set of all products that one can form from four elements of some linear basis or,
alternatively, by studying associativity for higher products of a set of generating
elements in the algebra. Some first attempt to carry the conformal bootstrap into
this direction was recently made in \cite{Vieira:2020xfx}, though in a somewhat
restricted setting.

From these short comments we can conclude that it is of significant interest to develop
a theory of multipoint conformal partial waves for $N > 4$ insertion points. The following
exposition of the subject in the context of integrability is based on several recent
publications and work in progress \cite{Buric:2020dyz,Buric:2021ywo,Buric:2021ttm}. There
have also been other activities in physics to study multipoint partial waves, in
particular directly through their integral representations. The most recent list of
references to this research can be found in \cite{Buric:2021ttm}.

To set the stage, let us state the well-known fact that the number of independent conformal
invariants one can build from $N$ points on $\mathbb{R}^d$ is given by
\begin{equation}
n_\textit{cr} (N,d) = \left\{ \begin{array}{ll} \frac12 N (N-3) \quad & N \leq d + 2 \\[3mm]
Nd - \frac12(d+2)(d+1) \quad &  N > d + 2 \end{array} \right.
\end{equation}
Following the approach we have described above, a complete characterization of the
associated conformal partial waves requires the same number of independent commuting
differential operators. As in the case of the 4-point function, the set of differential
operators depends on the choice of an OPE channel, see below for a precise definition.

\subsection{Gaudin integrable models}

The main idea for how to construct a complete set of commuting differential operators for
each OPE channel of an N-point function exploits an interesting relation with Gaudin
integrable systems. In order to make precise statements we need a bit of background on
Gaudin models~\cite{Gaudin_76a,Gaudin_book83}. Let us begin with a central
object, the so-called Lax matrix,
\begin{equation}
 \mathcal{L}(w)  = \sum_{i=1}^N \frac{\mathcal{T}^{(i)}_\a T^\a}{w-w_i}
 = \mathcal{L}_\a(w) T^\a\ .
\label{eq:Lax}
\end{equation}
Here, $w_i$ are a set of complex numbers, $T_\a$ denotes a basis of generators of the
conformal Lie algebra in $d$ dimensions and $T^\a$ its dual basis with respect to an
invariant bilinear form. The object $\mathcal{T}^{(i)}_\a$ denotes the set of right
invariant vector fields on the space $\Gamma^{\Delta_i}_{G/P}$.\footnote{Our
discussion is tailored to the study of N-point functions of scalar fields. To
accomodate tensor fields one needs to extend to more general flag manifolds.}

Given some conformally invariant symmetric tensor $\kappa_p$ of degree $p$ one can
construct a family $\mathcal{H}_p(w)$  of commuting operators as~\cite{Feigin:1994in,
Talalaev:2004qi,Molev:2013}
\begin{equation}\label{eq:HpGaudin}
\H_p(w) = \kappa^{\a_1 \cdots \a_p}_p \mathcal{L}_{\a_1}(w) \cdots
\mathcal{L}_{\a_p}(w) + \dots  \ ,
\end{equation}
where the dots represent correction terms expressible as lower degree combinations of
the Lax matrix components $\mathcal{L}_\a(w)$ and their derivatives with respect to $w$.
For $p=2$ such correction terms are absent. The correction terms are necessary to ensure
that the families commute,
\begin{equation}\label{eq:com}
[\, \H_p(w) \, , \, \H_q(w') \, ] = 0\ ,
\end{equation}
for all $p,q$ and all $w,w' \in \mathbb{C}$.  It is a well-known fact that these families
commute with the diagonal action of the conformal algebra, i.e.
\begin{equation}
[\, \mathcal{T}_\a\,  ,\,  \H_p(w)\, ] = 0 \quad \textit{ where }
\ \mathcal{T}_\a = \sum_{i=1}^N \mathcal{T}_\a^{(i)}\ .
\end{equation}
Hence the commuting families $\H_p(w)$ of operators descend to differential operators on
functions of the $n_{\textit{cr}}(N,d)$ conformally invariant cross ratios.

\subsection{OPE limits of the Gaudin model}

At this point, the commuting families $\H_p(w)$ still contain $N$ continuous parameters
$w_i, i= 1, \dots, N$. Without loss of generality we can set three of these complex numbers
to some specific values, e.g. $w_1 = 0, w_{N-1}=1, w_{N} = \infty$ so that we remain with
$N-3$ complex parameters our Gaudin Hamiltonians depend on. The set of differential
equations which characterize conformal partial waves, on the other hand, should only
depend on the discrete choice of an OPE channel. As we shall see now, any OPE channel
determines a certain limit of the parameters $w_i$. In other words, the parameters
$w_i$ are entirely fixed by the channel.

\begin{figure}
\begin{center}
\includegraphics[scale=1]{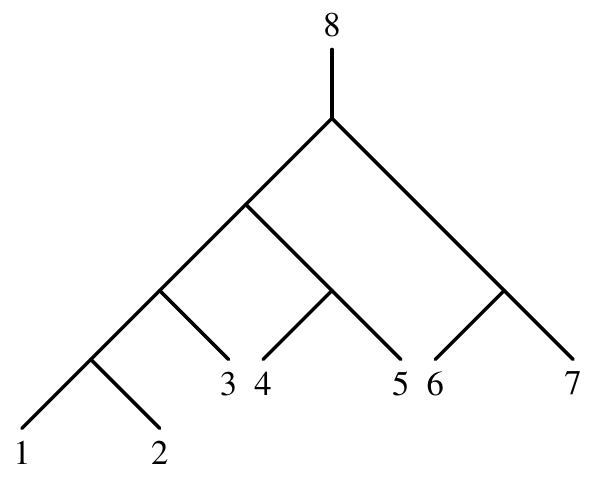}
\caption{Plane rooted tree representation of an OPE diagram with 8 external fields.}\label{fig:diag}
\end{center}
\end{figure}

In order to make a precise statement, we need a bit of preparation. Let us pick some OPE
channel $\mathcal{C}$ along with an (arbitrary) external edge in the diagram. The latter
serves as a reference point and we shall assign it the label $N$. As this edge is external,
it is attached to a unique vertex which we will denote by $\rho_\ast$. Such a choice of
reference vertex defines a so-called rooted tree representation of the channel, see
Figure \ref{fig:diag} for a concrete example with $N=8$. We draw the OPE diagram on
a plane, with the vertex $\rho_\ast$ situated at the top and with each vertex having two
downward edges attached. The representation on a plane forces us to make a choice of which
edges are pointing towards the left and which edges are pointing towards the right: this
choice is arbitrary, and gives rise to what is called a plane (or ordered) representation
of the underlying rooted tree.

Obviously, each vertex $\rho \in V$ from the set $V$ of $N-2$ vertices of the OPE diagram
defines a partition $\underline{N} = I_{\rho,1} \cupdot I_{\rho,2} \cupdot I_{\rho,3}$, with
the sets $I_{\rho,j}$ formed by the labels of the external fields attached to the three
branches of the vertex. We shall fix the labelling of the three branches using the plane
rooted tree representation of the channel: choose the branch $I_{\rho,1}$ to be the one
pointing to the bottom left while the branch $I_{\rho,2}$ is pointing the bottom right.
By construction, the last branch $I_{\rho,3}$ then always points to the top and contains
the reference point $N$. Let us also note that the vertices $\rho \neq \rho^*$ are
associated with a unique internal edge, namely the one that points upward. Thereby the
plane rooted tree representation allows us to identify the set $E$ of internal edges
with $E \equiv V\setminus \{\rho^* \}$.
\medskip

So far, vertices were just enumerated by an element $\rho \in V$. Once we have drawn the
diagram in a plane, we can assign a sequence $s^\rho =(s^\rho_1,s^\rho_2, \dots,
s^\rho_{\delta_\rho})$ of elements $s^\rho_a \in \{1,2\}$ to each vertex $\rho$. This sequence
$s^\rho$ encodes the path from $\rho_\ast$ to $\rho$. It tells us whether we have to move
to the left (for $s_a^\rho=1$) or right (for $s_a^\rho=2$) every time we reach a new vertex
until we arrive at $\rho$ after $\delta_\rho$ steps. We shall also refer to the length
$\delta_\rho$ of the sequence as the depth of the vertex and to $s^\rho$ as the
\textit{binary sequence} of $\rho$. Note that the top vertex $\rho_\ast$ has depth
$\delta_{\rho_\ast}=0$.

In order to construct the limit of the Gaudin model associated with the OPE channel
$\mathcal{C}$, we need now construct a polynomial $g_\rho(\varpi)$ for each vertex
$\rho$. If $s^\rho$ is the binary sequence associated with the vertex $\rho$, the
polynomial $g_\rho$ is defined as
\begin{equation} \label{eq:grho}
g_\rho(\varpi) = \sum_{a=1}^{\delta_\rho} \varpi^{a-1} \delta_{s_a^\rho,2} \ .
\end{equation}
Obviously the top vertex $\rho_\ast$ is assigned to $g_{\rho_\ast}(\varpi) = 0$. The vertices
of depth $\delta_\rho = 1$ are associated with $g_{\rho_1}(\varpi)= 0$ or $g_{\rho_2}(\varpi)
= 1$, depending on whether they are reached from $\rho_\ast$ by going down to the left
($\rho_1$) or to the right ($\rho_2$) etc.

Similarly, we can assign polynomials $f_i(\varpi)$ to each external edge $1 \leq i<N$ at
the bottom of the plane rooted tree. Once again, we can encode the path from $\rho_\ast$
down to the edge $i$ by a binary sequence $s^i = (s^i_1, s^i_2,  \dots, s_{\delta_i}^i)$.
The length $\delta_i$ of the sequence $s^i$ is also referred to as the depth of the edge
$i$. Now we introduce
\begin{equation} \label{eq:fi}
f_i(\varpi) = \sum_{a=1}^{\delta_i} \varpi^{a-1} \delta_{s^i_a,2}
+ \varpi^{\delta_i}\delta_{s^i_{\delta_i},1}\ .
\end{equation}
and set
\begin{equation}  \label{eq:wHighest}
f_N(\varpi) = \varpi^{-1}
\end{equation}
for the external edge of the reference field at the top of the plane
rooted tree. We have thus set up all the necessary notation that is needed to
construct the relevant scaling limits of the $N$-site Gaudin model that remove the
dependence on the continuous parameters $w_i$
\medskip

The main idea in defining the channel dependent limit is to set $w_i = f_i(\varpi)$
and to send $\varpi$ to zero. When we perform this limit of the objects
$\mathcal{H}^{(p)} (w;w_i)$ of the $N$-site Gaudin model we must adjust the
spectral parameter $w$ and multiply with some appropriate powers of $\varpi$ in
order to obtain sufficiently many independent commuting operator. More precisely,
for given a vertex $\rho$ we introduce
\begin{equation}\label{eq:LimitH}
\Hc_\rho^{(p)}(w) = \lim_{\varpi \to 0} \varpi^{pn_\rho}
\Hc^{(p)} \bigl( \varpi^{n_\rho}w + g_\rho(\varpi),w_i = f_i(\varpi) \bigr)\ .
\end{equation}
It can be shown that the set of these objects with $\rho \in E$ indeed contain
$n_{\textit{cr}}(N,d)$ independent operators on the space of conformal invariants.
These include operators of the form
\begin{equation}
\mathcal{C}^{(p)}_\rho = \lim_{w \to \infty} w^p \Hc_\rho^{(p)}(w)  \quad
\mathit{ for } \quad \rho \neq \rho^*\ .
\end{equation}
These measure the weight and spin of the field which is exchanged along the internal
edge $\rho \in E$ above the vertex $\rho \neq \rho^*$. We refer to the
operators $\mathcal{C}^{(p)}_\rho$ as Casimir operators of the edge $\rho$.  For a
given edge $\rho \neq \rho^*$ there can be relations between operators with different
values of $p$. Taking these into account, the internal edge $\rho \neq \rho^*$
provides us with
\begin{equation} \label{eq:deldef}
n_{\textit{Cas}}(\rho,d) :=
\textrm{min}\left(|I_{\rho,3}|, N - |I_{\rho,3}|, \textrm{rank}_d\right)
=: \dep_\rho
\end{equation}
independent operators. Here $|I|$ denotes the order of the subset $I \subset
\underline{N}$ and $\mathrm{rank}_d$ is the rank of the $d$-dimensional conformal
algebra. It is easy to see that for $N > 4$, the integers $n_{\textit{Cas}}$ do not
add up to $n_{\textit{cr}}$. This means that the Casimir operators do not suffice
to characterize multipoint conformal partial waves. The missing operators can all
be constructed from the objects \eqref{eq:LimitH}. It turns out that given a vertex
$\rho$ and an index $p$ one can prepare
\begin{equation}
    n_{\textit{vert}}^{(p)}(\rho,d)
    = \textit{max}\left[\left((p-2)-\sum_{e \in \overline{E}_\rho}^3
    \Theta_0(p-2\dep_e)(p-2\dep_e-1)\right),0\right]\,,
    \label{eq:countingnormal}
\end{equation}
additional independent \textit{vertex differential operators}. Here $\Theta_0$ is
the Heaviside step function with $\Theta_0(0)=0$ and $\overline{E}_\rho$ consists
of all external and internal edges that have one endpoint at the vertex $\rho$. For
internal edges $e = \rho'$, the quantity $\dep_e$ was introduced as a shorthand
in eq.\ \eqref{eq:deldef} and for external edges we set $\dep_e = 1$. The formula
applies for all $p$ as long as $d$ is odd. It requires some adjustment in even $d$,
see \cite{Buric:2021ttm} for details and a derivation.

Given this counting of independent commuting differential operators it is now
possible to check that
\begin{equation}
n_\textit{cr}(N,d) = \sum_{\rho \neq \rho^*} n_{\textit{Cas}}(\rho,d) +
\sum_{\rho} \sum_p n_{\textit{vert}}^{(p)}(\rho,d)\ ,
\end{equation}
i.e. that the set of independent differential operators suffices to completely
characterize the multipoint conformal partial waves.

\begin{figure}[H]
\begin{center}
\includegraphics[scale=1]{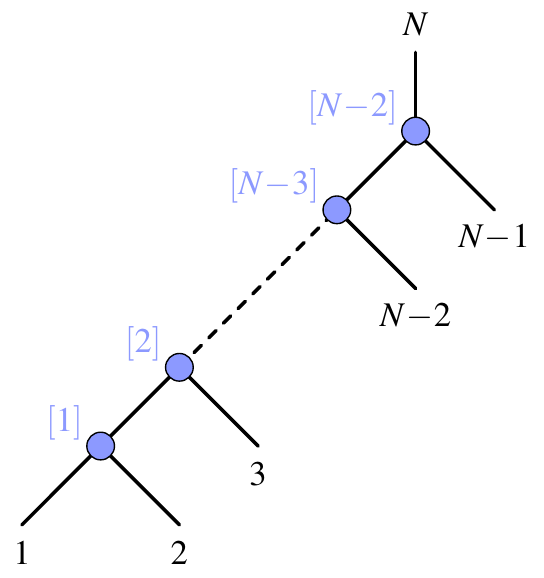}
\caption{Choice of plane rooted tree representation of the comb channel OPE diagram
with $N$ points.}\label{fig:comb}
\end{center}
\end{figure}

The OPE limits we described here are generalizations of the so-called caterpillar
or bending flow limits that have been considered in the mathematical literature,
see e.g. \cite{Chervov:2007dn,Chervov:2009,rybnikov2016cactus}. The caterpillar limit
corresponds to the special case of an OPE limit for which the OPE diagram has the
shape of a comb, see Figure \ref{fig:comb}. Following our conventions in
\cite{Buric:2020dyz}, we enumerate the vertices $\rho = [r]$ by an integer $r = 1,
\dots, N-2$ that is related to its depth $\delta_\rho = \delta_{[r]}$ as $\delta_{[r]}
= N-2-r$.
One can compute the limit of the Gaudin model associated with this tree using the
construction outlined in the previous subsection. For the polynomials $f_i$ that
determine the parameters $w_i$ of the Gaudin model, one finds from eq. \eqref{eq:fi}
that
\begin{equation}
w_i = f_i(\varpi) = \varpi^{N-1-i}, \qquad \forall\,i\in\lbrace 1,\cdots,N \rbrace\ .
\end{equation}
while all the vertex polynomials vanish, i.e. $g_{[r]}=0$ for $r=1, \dots, N-2$.
Our general prescription \eqref{eq:LimitH} thus becomes
\begin{equation}
\Hc_{[r]}^{(p)}(w) = \lim_{\varpi \to 0}
\varpi^{p(N-2-r)} \Hc^{(p)} \bigl( \varpi^{N-2-r} w;w_i = \varpi^{N-1-i}\bigr)\ .
\end{equation}
In order to illustrate our results and also prepare for the discussion in the next
subsection we want to take a closer look at the case of $N=6$ points for $d > 4$.
According to equation \eqref{eq:deldef}, the three internal edges $[r], r = 1,2,3,$
in the comb channel diagram contribute
\begin{equation} \label{eq:ncasex}
n_{\textit{Cas}}([1],d) = 2 = n_{\textit{Cas}}([3],d) \quad , \quad
n_{\textit{Cas}}([2],d) = 3
\end{equation}
and hence a total number of $7$ Casimir differential operators. Our result
\eqref{eq:countingnormal}, on the other hand shows that the four vertices
$[r], r=1,\dots, 4,$ contribute
\begin{equation}
n_{\textit{vert}}^{(p)}([1],d) = 0 = n_{\textit{vert}}^{(p)}([4],d)
\quad, \quad  n_{\textit{vert}}^{(p)}([2],d) = 1 = n_{\textit{vert}}^{(p)}([3],d)
\end{equation}
and hence a total of two vertex differential operators which are both of order
four. When combined, the number of differential operators matches the number of
$n_{\textit{cr}} = 9$ cross ratios.

\subsection{Relation with lemniscatic elliptic CMS model}

The two inner vertices of the $N=6$ point partial wave are among the simplest
vertices that carry degrees of freedom, i.e. for which $n_\textit{vert} \neq 0$. As one
can infer from our general formula \eqref{eq:countingnormal}, it is in fact the
unique vertex in $d > 4$ that gives rise to a single parameter. Let us focus on
the vertex $\rho = [2]$ of the $N=6$ point function, see our discussion at the
end of the previous subsection, which connects the internal edges $[1]$ and
$[2]$ with the external line $i=3$. According to our list \eqref{eq:ncasex},
the edge $[2]$ admits two independent Casimir operators whose eigenvalues we
parametrize through $\Delta_1$ and $l_1$. The internal edge $[3]$, on the
other side, admits three independent Casimir operators. Their eigenvalues
can be parametrized by three parameters $\Delta_2, l_2, \ell_2$. The external
scalar edge comes with one parameter, namely the scaling weight $\Delta_3$ of
the external scalar field.\footnote{This is different from the weight on the
third internal edge which we also denoted by $\Delta_3$ in the previous
subsection.} Let us note that for $d=4$, the vertex which we
consider here is the special case of a single variable vertex with one additional
parameter $\ell_1 \neq 0$. On the other hand, if we specialize the non-trivial
vertex of the $N=6$ point function to $\ell_2 = 0$ it can be considered down
to $d=3$.
\begin{figure}
\begin{center}
\includegraphics[scale=.5]{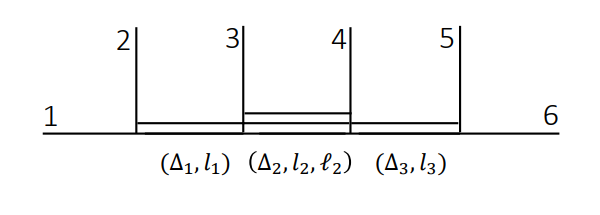}
\caption{The two nontrivial vertices in the 6-point function are
characterized by six quantum numbers each.}\label{fig:6pt}
\end{center}
\end{figure}

Our goal now is to identify the fourth order vertex differential operator for
the single variable vertices. The easiest way to describe the findings of
\cite{Buric:2021ttm} involves a new algebra that is generated by operators
$A,A^{\dagger}$ and $N$ which obey the following
relations
\begin{eqnarray}
    [N,A^{\dagger}]  =   A^{\dagger} \quad  & , & \quad [N,A]=-A\,, \label{NA*NA} \\[2mm]
     A A^{\dagger}  = R_\ag(N+1) \quad & , & \quad    A^{\dagger} A = R_\ag(N) \label{AA*A*A}
\end{eqnarray}
where
$$
R_\ag(N) = \frac{N(N+2\ag-1)}{(N+\ag-1)(N+\ag)} (N+\nu_1+2\ag)
    (N-\nu_1-1)(N+\nu_2+2\ag)(N-\nu_2-1)\,.
$$
The algebra depends parametrically on the three spin parameters as $\ag = \ell_2 +(d-3)/2$
and $\nu_k = l_k - \ell_2$, but is independent of the conformal weights $\Delta_i$. The
family of algebras comes equipped with an involutive anti-automorphism $\ast$ defined by
$N^\ast = N$ and $A^\ast = A^\dagger$, provided that the parameters obey
\begin{equation}
    \bar{\ag} = \ag\,, \qquad \bar{\nu}_1 = -(2 \ag+1+\nu_1)\,, \qquad
    \bar{\nu}_2 = -(2 \ag+1+\nu_2)\,.
\end{equation}
The vertex operator $H$ can now be written as a rational combination of the three
generators as,
\begin{align}
H^{(d;\Dg_i;l_i;\ell_i)} =& B^{\dagger} B - \Gamma  (N+\ag)^2+
\frac{\ag(\ag-1)\, K}{(N+\ag)^2-1} + E^{(d;\Dg_i;l_k;\ell_2)}.
\label{HamFac}
\end{align}
Here, we defined the operators
\begin{align}
B^{\dagger} :=& \frac{A-A^{\dagger}}{2\mathrm{i} } - \mathrm{i} (2\cg_1\cg_2 +
\mathrm{i} \cg_3)(N+\ag)\,,  \\
B :=& \frac{A-A^{\dagger}}{2\mathrm{i} } + \mathrm{i} (2\cg_1\cg_2 -\mathrm{i}
\cg_3)(N+\ag)\,,
\end{align}
and the two parameters
\begin{align}
\Gamma :=& \frac{1}{4} (1+4\cg_1^2)(1+4\cg_2^2)\,, \\
K :=&  (\nu_1+\ag)(\nu_2+\ag)(\nu_1+\ag+1)(\nu_2+\ag+1)\,.
\end{align}
An expression for the constant term $E$ in our formula for the vertex operator
\eqref{HamFac} can be found in subsection 5.2 of \cite{Buric:2021ywo}. The three
parameters $\gamma_i$ that appear in the last few formulas are related to the
conformal weights $\Delta_i$ of the three fields at the vertex as
$$ \Delta_i = \frac{d}{2} + \mathrm{i} \gamma_i \ . $$
The operator $H$ is self-adjoint provided that the parameters $\gamma_i$ are real.

As one can easily check, the algebra generated by $A, A^\dagger$ and $N$ admits a
representation in a basis $C_n^{(\ag)}$ of the form
\begin{align}
    N C_n^{(\ag)} &:= n C_n^{(\ag)}, \label{NCn} \\
    A C_n^{(\ag)} &:= (n+ \nu_1+2\ag)(n+\nu_2+2\ag)
    \frac{n+2\ag-1}{n+\ag} C_{n-1}^{(\ag)}\,,\label{ACn} \\
    A^{\dagger} C_n^{(\ag)} &:= (n-\nu_1)(n-\nu_2) \frac{n+1}{n+\ag}
    C_{n+1}^{(\ag)}, \label{A*Cn}
\end{align}
where eq.\ \eqref{ACn} applies to all integers $n>0$, and $A C_0^{(\ag)} = 0$ when
$n=0$, i.e.\ the state $C_0^{(\ag)}$ is annihilated by the lowering operator $A$.
Similarly, the action of the raising operator $A^\dagger$ vanishes if $n=\nu_1$ or
$n=\nu_2$. Consequently, one can restrict the action of $A,A^\dagger$ and $N$ to
the finite dimensional subspace that is spanned by $C_n$ for $n = 0,\dots,
\mathrm{min} (\nu_1,\nu_2)$. When the basis is realized as Gegenbauer polynomials
$C^{(\ag)}(s) = C^{(\ag)}(1-2\mathcal{X})$, the operators $A,A^{\dagger}$ describe
the action of second order differential operators in the variable $\mathcal{X}$. We
then showed that the resulting fourth order differential operator $H= H_\mathcal{X}$
can be mapped to the Hamiltonian $L$ of a crystallographic elliptic Calogero-Moser-%
Sutherland model that was discovered originally by Etingof, Felder, Ma and Veselov
\cite{etingof2011107}. The latter acts on functions in a variable $z$ on the
lemniscatic elliptic curve. We found the variables $\mathcal{X}$ to be related
to $z$ as
\begin{equation}
\mathcal{X} = \frac{\wp(1/2)^2}{\wp(1/2)^2- \wp(z)^2}\ ,
\label{X_z_CoV}
\end{equation}
where $\wp$ is the Weierstrass elliptic function with lemniscatic periodicity conditions
$\wp(z) = \wp(z+1) = \wp(z+i)$. Given this change of variables, our vertex Hamiltonian
$H = H_\mathcal{X}$ and the elliptic CMS Hamiltonian $L = L_z$ of \cite{etingof2011107}
are related as
\begin{equation}
  \Theta^{-1} H_{\mathcal{X}} \Theta  \sim  L_z
\end{equation}
where
\begin{equation}
    \Theta \sim \, \mathcal{X}^{(l_1+l_2-2\ell_2 +
    \Delta_3 + (1-d)/2)/4} (1-\mathcal{X})^{(l_1+l_2-2\ell_2-
    \Delta_3+(1+d)/2)/4}\ .
\end{equation}
The lemniscatic Calogero-Moser-Sutherland Hamiltonian $L$ involves seven parameters
(multiplicities) whose values are determined by the three conformal weights $\Delta_i$,
the spin parameters $l_1, l_2, \ell_2$ and the dimension $d$. Precise formulas can be
found in \cite{Buric:2021ttm}. A similar relation for a 3-point vertex with one scalar
and two generic spinning legs of the $d=4$ dimensional conformal group also holds. In
that case, the dimension $d$ is fixed and the seven multiplicities of the lemniscatic
model $L$ are obtained from the three conformal weights and four spin variables $l_1,
\ell_1,l_2,\ell_2$, see \cite{Buric:2021ttm} for precise formulas.

\section{Concluding comments}

This brings us to the end of this short note on higher dimensional CFT and its
profound relation to multivariate hypergeometry. In the past, the interaction
between these two fields has not lived up to its true potential. I hope that
this text can make a modest contribution to developing a culture of mutual
interest and understanding. It would clearly be very beneficial to join forces
on overcoming some of the challenges that arise within CFT.

In my outline of concrete open problems I have focused on the study of multipoint
partial waves, mostly because it is particularly timely and has seen some recent
advances that are partly due to the very fruitful interactions I had during and
after the workshop. Let me stress, however, that there are many other interesting
problems that could profit from fresh mathematical input. One of these concerns
the study of hypergeometric functions associated with (conformal) supergroups. I
have only scratched this topic very briefly at the end of section 3. Given that
most of the CFTs we know about are supersymmetric, progress in the theory of
superconformal partial waves is guaranteed to have many applications. While
physicists have made some advances over the last years, very much remains to be
done.


\bibliographystyle{amsplain}
\providecommand{\bysame}{\leavevmode\hbox to3em{\hrulefill}\thinspace}
\providecommand{\MR}{\relax\ifhmode\unskip\space\fi MR }
\providecommand{\MRhref}[2]{%
  \href{http://www.ams.org/mathscinet-getitem?mr=#1}{#2}
}
\providecommand{\href}[2]{#2}


\end{document}